\documentclass[10pt,a4paper]{article}

\begin{document}

\large
\title{Pressure-tuning of the electron-phonon coupling: the insulator to metal transition in manganites.}
\author{P. Postorino, A. Congeduti, P. Dore, F.A. Gorelli, \and L. Ulivi, A. Sacchetti, A. Kumar, D.D. Sarma}
\maketitle

\begin{abstract}
A comprehensive understanding of the physical origin of the unique 
magnetic and transport properties of $A_{1 - 
x}A'_{x}$MnO$_{3}$ manganites ($A$ = trivalent 
rare-earth and $A'$ = divalent alkali-earth metal) is still far from being 
achieved \cite{1,2,3}. The complexity of these systems arises from 
the interplay among several competing interactions of comparable strength. 
Recently the electron-phonon coupling, triggered by a Jahn-Teller distortion 
of the MnO$_{6}$octahedra, has been recognised to play an 
essential role in the insulator to metal transition and in the closely 
related colossal magneto-resistance \cite{3}. The pressure tuning of 
the octahedral distortion gives a unique possibility to separate the basic 
interactions and, at least in principle, to follow the progressive 
transformation of a manganite from an intermediate towards a weak 
electron-phonon coupling regime. Using a diamond anvil cell, temperature and 
pressure-dependent infrared absorption spectra of 
La$_{0.75}$Ca$_{0.25}$MnO$_{3}$ have been 
collected and, from the spectral weight analysis \cite{4}, the 
pressure dependence of the insulator to metal transition temperature 
$T_{IM}$ has been determined for the first time up to 11.2 GPa. The 
$T_{IM}(P)$ curve we proposed to model the present data revealed a 
universality character in accounting for the whole class of intermediate 
coupling compounds. This property can be exploited to distinguish the 
intermediate from the weak coupling compounds pointing out the fundamental 
differences between the two coupling regimes.
\end{abstract}

The crystal structure of manganites consists of a pseudocubic lattice of 
MnO$_{6}$ octahedra with the $A-A'$ ions placed in the free volume among them. 
In doped compounds (0 $< x <$ 1), the Mn ions assume the Mn$^{ + 3}$/Mn$^{ + 4}$ 
mixed-valence and the Mn$^{ + 3}$O$_{6}$ octahedra undergo an asymmetric 
distortion which removes the degeneracy of the outer Mn d-orbital with 
$e_{g}$ symmetry (Jahn-Teller (JT) effect). The presence of distorted 
octahedra in the pseudocubic lattice gives rise to a local potential well 
which, in turn, tends to localise the $e_{g}$-electrons. This self-trapping 
mechanism is a fingerprint of the electron-phonon coupling (EPC) whose 
strength depends on the extent of JT distortion. Therefore, tuning the 
asymmetry of the Mn$^{ + 3}$O$_{6}$ by means of the so-called \textit{internal pressure} (chemical 
substitution which changes the average $A-A'$ ionic radius $R_{A})$ \cite{5,6,7} 
or by an external pressure (hydrostatic pressure P) \cite{8,9,10,11,12,13,14,15,16,17}, is an 
intriguing way to probe the EPC and its localising tendency.

EPC profoundly affects the transport properties of manganites, which range 
from insulating (strong EPC) to metallic (weak EPC) character \cite{18}. At 
intermediate-coupling (IC), several manganites (0.2$< x <$0.5) show Colossal 
Magneto-Resistance (CMR) and, on cooling, they undergo a first-order 
insulator-to-metal (IM) transition, whereas Weak-Coupling (WC) systems show 
moderate magneto resistance and a continuous IM transition \cite{18,19,20}. 
Since the temperature $T_{IM}$ is expected to be proportional to the 
effective $e_{g}$ orbital bandwidth $W_{eff}$ \cite{6,8}, shifts in $T_{IM}$ 
can be produced by forcing $W_{eff}$ to change. This can be achieved by a 
pressure-tuning of the EPC, which behaves as a narrowing factor $\xi  \le 
$1 applied on the bare structural term $W_{0}$ \cite{7}, that is $T_{IM} 
\propto W_{eff}=W_{0}\xi$ \cite{8,21}. \textit{Internal} or external applied pressure 
cause an increase of $T_{IM}$ through the double mechanism of enlarging 
$W_{0 }$ by enhancement of the Mn-O bond covalency, and of increasing $\xi $ 
by a reduction of the EPC strength. Exploiting chemical substitution, 
$R_{A}$-$T$ phase diagrams have been obtained \cite{5,22}, whereas a set of 
$T_{IM}(P)$ data \cite{8,9,10,11,12,13,14,15,16,17} is available up to a maximum of 2 GPa, owing to 
the many difficulties of high-pressure transport and magnetic measurements. 
The experimental data suggest an equivalence between \textit{internal} and external 
pressure \cite{5}: the same change of $T_{IM}$ can be achieved by varying 
either the average ionic radius by $\Delta R_{A}$ or the pressure by 
$\Delta P$=$\Delta R_{A}/ \beta$ ($\beta $=3.7 10$^{ - 3}$ \AA /GPa 
\cite{5}). Nevertheless, external pressure has the clear advantage of 
operating on a single sample in a continuous way and under conditions more 
controlled than chemical substitution.

The IM transition can be conveniently described in terms of polarons, 
quasi-particles associated to a charge plus the surrounding lattice 
distortion. Indeed, a picture of localized polarons is associated to the 
paramagnetic-insulating phase of the manganites, whereas in the 
ferromagnetic-metallic phase below $T_{IM}$, the increase of the 
$e_{g}$-electron kinetic energy, due to the magnetic double-exchange 
interaction, favors polarons itinerancy \cite{4,23,24}. Since the transition 
of polarons from localised states to the continuum gives rise to a large 
band centred in the near Infrared (IR)\cite{4,24} the delocalization process 
can be monitored by spectroscopic techniques. Indeed, the band-profile 
reflects the distribution of the polaron binding energies and the spectral 
weight, that is the frequency-integrated absorption spectrum, provides a 
measure of the $e_{g}$-electron kinetic energy and, hence, of the polaron 
mobility \cite{4,24,25}. Exploiting the advantage of coupling of Diamond Anvil 
Cells (DAC) to IR spectroscopy \cite{4}, the delocalization process can be 
monitored over an extended and, until now, unexplored pressure region. 

La$_{0.75}$Ca$_{0.25}$MnO$_{3}$ ($T_{IM}(P$=0$)$=220 K \cite{4}) was chosen as a 
suitable IC candidate for the present experiment and temperature-dependent 
IR spectra (100-320 K) were collected along several isobaric paths (0 $<P<$ 11.2 
GPa) over the 500-4500 cm$^{ - 1 }$spectral range. The DAC equipped with 600 
$\mu $m culet IIA diamonds and steel gasket (300-$\mu $m diameter and 
50-$\mu $m deep hole) was mounted on the cold finger of a cryocooler. The 
Bruker 120HR interferometer, at LENS (European Laboratory for Non-Linear 
Spectroscopy, Florence), was equipped with a KBr beam-splitter and an MCT 
detector. High-quality measurements, in spite of the high absorbance and the 
small size of the sample, were obtained thanks to the high-efficiency 
focusing system \cite{26}. The sample, prepared by a solid state reaction 
method \cite{27}, was finely milled and smeared on the top surface of a KBr 
pellet sintered inside the gasket hole \cite{4}. The optical density $Od(\nu 
)=\ln [I_{0}(\nu )/I(\nu )]$ was obtained by measuring the intensities 
$I_{0}(\nu )$ and $I(\nu )$, transmitted by the pure KBr pellet and by the 
KBr plus sample \cite{4}. The $Od(\nu )$ measured along three representative 
isobaric paths at selected temperatures are shown in Fig. 1. The intense 
peak around 600 cm$^{ - 1}$ is due to the IR active $B_{2u}$ phonon mode 
\cite{4}, while the broader structures around 1000 cm$^{ - 1}$ and 1400 cm$^{ 
- 1}$ originate from multi-phonon processes, enhanced by resonance 
effects \cite{28}. The phonon nature of these features is confirmed by their 
moderate pressure-induced frequency hardening \cite{4,29}. The phonon spectrum 
is superimposed to a broad and intense contribution, which extends over the 
whole spectral range, is strongly temperature and pressure dependent and 
arises from the low frequency side of the polaronic band. Upon cooling at a 
given pressure, the overall absorption decreases (red spectra) until an 
\textit{inversion temperature} is reached, where the trend is reversed and 
the absorption starts to increase (blue spectra). The \textit{inversion temperature} 
identifies $T_{IM}$ \cite{4}, since the observed absorption trend corresponds 
to what expected for the mobility of the charge 
carriers in the insulating or metallic phase. The temperature dependence of 
the spectral weights $n^{\ast}(T)$ at different pressures was obtained by 
integrating the measured $Od(\nu )$ over the 800-1800 cm$^{ - 1}$ range. In 
Fig. 2 $n^{\ast}(T)$ are shown by red and blue symbols to distinguish the insulating 
($dn^{\ast}/dT<$0) from the metallic ($dn^{\ast}/dT > 0$) regime \cite{4,18}. 
Fig. 2 shows that at fixed temperature $dn^{\ast}/dP>$0, 
meaning that pressure causes an increase of polaron mobility over 
the whole temperature range. At low-temperatures ($T<$150K), all the $n^{\ast}(T)$ 
seem to converge towards the same asymptotic curve, suggesting the occurrence 
of a sort of \textit{asymptotic metal }state, consistent with the picture of fully itinerant polarons.

The fine temperature sampling of $n^{\ast}(T)$ enables an accurate determination of 
the $T_{IM}(P)$ values, shown in the P-T phase diagram of Fig. 3a. The 
deviation from linearity, apparent in the pressure-dependence of $T_{IM}$, 
is a new and unexpected finding, unpredicted by the simple linear 
extrapolation of the low-pressure data available in literature. The present 
data are successfully described by the empirical curve

\begin{equation}
T_{IM}(P) = T_{\infty }-[T_{\infty }-T_{IM}(0)]exp(-P/ P^{\ast })
\label{eq1}
\end{equation}

\noindent
with $T_{\infty }$=299$\pm $3 K and P$^{\ast }$=3.4 $\pm $0.4 GPa. Using 
structural data at high pressure \cite{27}, we estimated the pressure 
dependence of $W_{0 }$ \cite{8} and, hence, the bare structural contribution to 
$T_{IM}$(P) (dashed line in Fig. 3a). It is apparent that the bare 
contribution alone does not account for the observed rapid rise of $T_{IM}$ 
at low-pressure, but a considerable pressure-driven weakening of the EPC, 
that is a considerable broadening of the effective bandwidth $W_{eff}$ (i.e. 
$\xi  \to $1), is necessary. On the contrary, the bare structural term 
adequately reproduces the low P-sensitivity in the high-pressure regime. 
These findings are also supported by the simple argument offered by the 
Clausius-Clapeyron equation, that is $dT_{IM}$/$dP$ = $\Delta v$/$\Delta s$. 
The rapid decrease of $dT_{IM}$/$dP$ vs $P$ should be more likely attributed to 
a reduction of the volume variation $\Delta v$ than to an increase of the 
entropy variation $\Delta s$. Indeed, applied pressure is expected to 
diminish the octahedral distortion in the insulating phase, thus reducing 
the volume discontinuity $\Delta v$ at the IM transition. The remarkable 
decrease of volume effects, recently reported for $R_{A}>$1.22 \AA \cite{19}, is 
also consistent with the vanishingly small values of $dT_{IM}$/$dP$ at high 
pressure (i.e. $\Delta v \to $0)

To state Eq. 1 in terms of a general law, we applied it to extract the 
dependence of $T_{IM}$ on $R_{A}$ at ambient pressure starting from a 
compound with a given $R_{A}^{0}$. No adjustable parameter was introduced 
and only the \textit{internal/}external pressure conversion was exploited through the variable 
substitution $P$ = $(R_{A} - R_{A}^{0})$/$\beta $. The $T_{IM}(R_{A})$ 
curve thus obtained is shown in Fig. 3b in comparison with the experimental 
data of several compounds at constant $x$=0.25 doping (Ref. 20). The agreement 
is excellent up to $R_{A} \sim $ 1.22 \AA, that is over the region 
corresponding to IC regime, and first order IM transition \cite{19,20}. At 
larger $R_{A}$ values (continuous transition \cite{19,20}, WC regime) the curve 
fails to describe the experimental data. The pressure derivative of Eq.\ref{eq1} 

\begin{equation}
\label{eq2}
\frac{dT_{IM}(P)}{dP} \Big| _{P}= \frac{T_{\infty }-T_{IM}(P)}{P^{\ast }}
\end{equation}

\noindent
is plotted versus $T_{IM}(P)$ in Fig. 4. Also shown in Fig. 4 are the 
zero-pressure limits of the experimental derivative $dT_{IM}(P)$/$dP|_{P = 
0}$ obtained from the present data (23 K/GPa) and from the $T_{IM}$(P) data 
available in literature for other manganites \cite{8,9,10,11,12,13,14,15,16,17}.
Also in this case, 
Eq. \ref{eq2} describes the IC regime rather well, at least for $T_{IM} >$200 K, 
whereas it is not appropriate for WC compounds (blue symbols). The deviation 
of the low $T_{IM}$ ($ \le $ 200 K) compounds from Eq.\ref{eq2} can be explained by 
the EPC getting stronger, with the localizing effect being strengthened by a 
high cation disorder \cite{22,30}. More quantitatively, for these samples the 
mean square deviation $\sigma _{cat}$ of the $A$-$A'$ ion radii is larger than 
the mean square thermal displacement $\sigma _{th}$, obtained from the 
experimental vibrational frequencies of the $A$-$A'$ ions. This results in an 
\textit{effective disorder} $\Sigma_{eff}$=$\sigma _{cat}$/$\sigma _{th} >$1. The universal 
behaviour here observed (Fig. 4) shows that, regardless of doping or 
chemical composition, $T_{IM}$ (i.e. $W_{eff}$) is actually the only relevant 
parameter for IC manganites. ($\Sigma_{eff}<$1).

The present study offers the opportunity of separating the contributions of 
the bare structural effects from the pressure-tuned EPC. The latter plays an 
essential role up to $P \approx $2$P^{\ast }$ ( $\approx$7 GPa), beyond 
which pressure-driven structural effects are sufficient to describe the IM 
transition curve (Fig. 3a). When reverted into an $R_{A}$-dependence, the 
pressure-dependence of $T_{IM}$, which shadows the EPC pressure-dependence, 
emphasises the full applicability of the model to IC systems (Fig. 3b). The 
departure of WC systems from the model curve suggests an abrupt readjustment 
of the balance among the different interactions. Indeed, when the 
internal/external pressure exceed a certain value (i.e. $R_{A} \to$ 1.22 
\AA or $P>$2$P^{\ast })$ their effect is no longer equivalent: while P leaves 
unchanged both the crystal symmetry and the EPC strength, $R_{A}$ induces a 
structural transition from orthorhombic to rhombohedral, whose higher 
symmetry does not allow for the static JT distortion \cite{19}. The transition 
from the IC to the WC regime is not continuous and cannot be achieved by 
simply tuning the EPC by means of external pressure. Eq. 1 enables to 
clearly distinguish between the two regimes and represents a strict 
benchmark for any theoretical model aimed at addressing the complex physics 
of these systems.

\newpage

{\bf Figure Captions.}

Fig. 1 $T$-dependent optical densities measured along three representative 
isobars. On cooling down, the overall absorption at first decreases (red 
curves, insulator) and then increases (blue curves, metal) as shown by the 
broken arrows. Data over 1850-2450 cm$^{ - 1}$ frequency range are not shown 
since the strong absorption from the diamonds affects the data quality.

Fig. 2 Temperature dependence of the spectral weight $n^{\ast}$ along seven isobaric 
paths. Red and blue symbols refer to insulating and metallic behaviour 
respectively, lines are a guide to the eye. 

Fig. 3 a) $P-T$ phase diagram of La$_{0.75}$Ca$_{0.25}$MnO$_{3}$. The grey 
line is the best-fit curve from Eq. 1 ($T_{\infty }$=299 K and $P^{\ast 
}$=3.4 GPa) to the present experimental data (half-filled circles). 

b) $R_{A}-T$ phase diagram for $x$=0.25 manganites: diamonds from Ref. 20, the 
grey line is obtained from Eq. 1 exploiting the internal-external pressure 
conversion with no adjustable parameters $T_{IM}(R_{A})$ = $T_{\infty 
}$-[$T_{\infty }$-$T_{IM}(R_{A}^{0})$]exp[-($R_{A}$- 
$R_{A}^{0})$/($\beta $/$P^{\ast })$]. To make the data internally 
comparable, the $R_{A}$ values of Ref. 20 have been recalculated using the 
ionic radii with coordination number 9 (Ref. 5) instead of 12 (ref. 20). The 
corresponding pressure scale is shown on the top x-axis.

Fig. 4 $T_{IM}$ pressure derivative vs. $T_{IM}$. Zero pressure limit 
($dT_{IM}(P)$/$dP|_{P = 0}$ vs. $T_{IM}$(0)): bold half-filled circle, 
present experiment; diamonds data from literature (the reference number is 
also shown). Light half-filled circles show $dT_{IM}(P)$/$dP|_{P}$ vs. 
$T_{IM}(P)$ from the present experiment. The grey line is from Eq. 2.

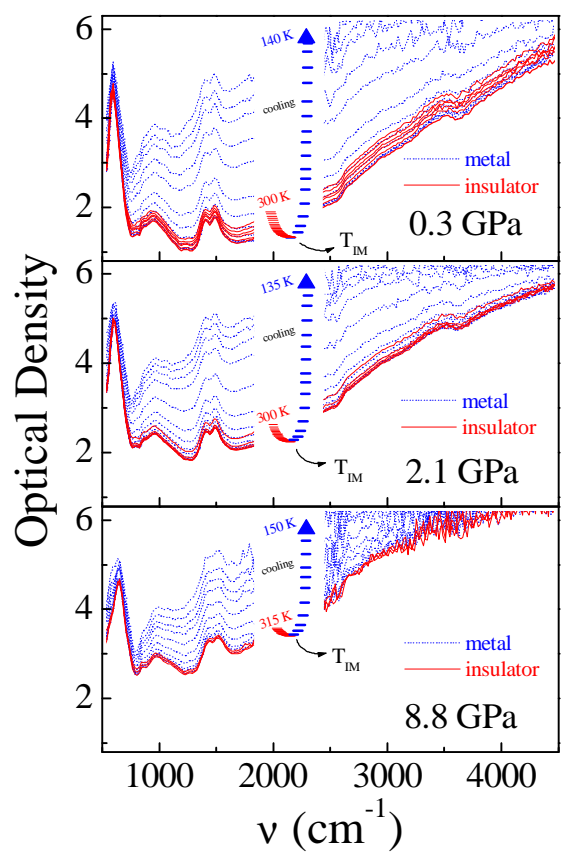
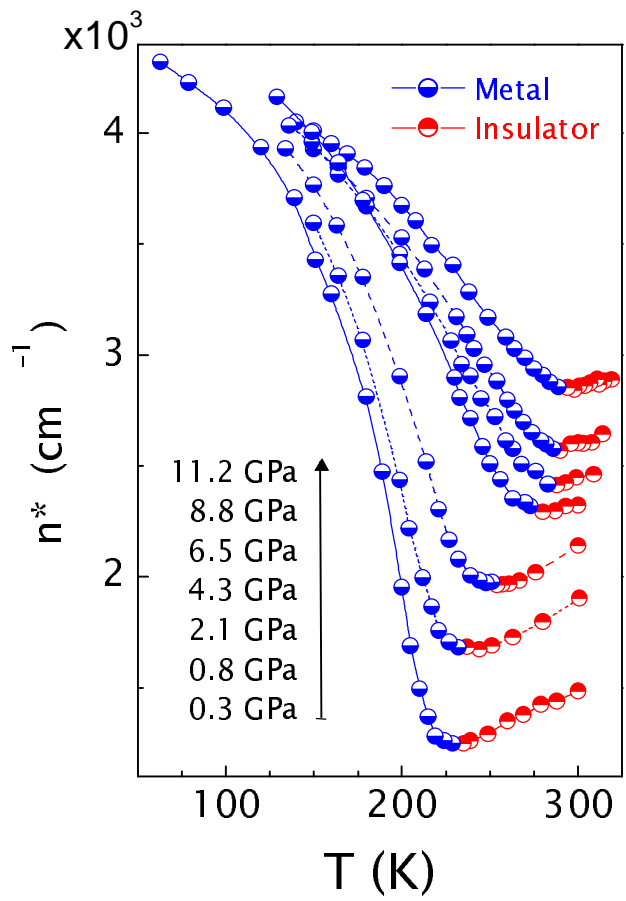
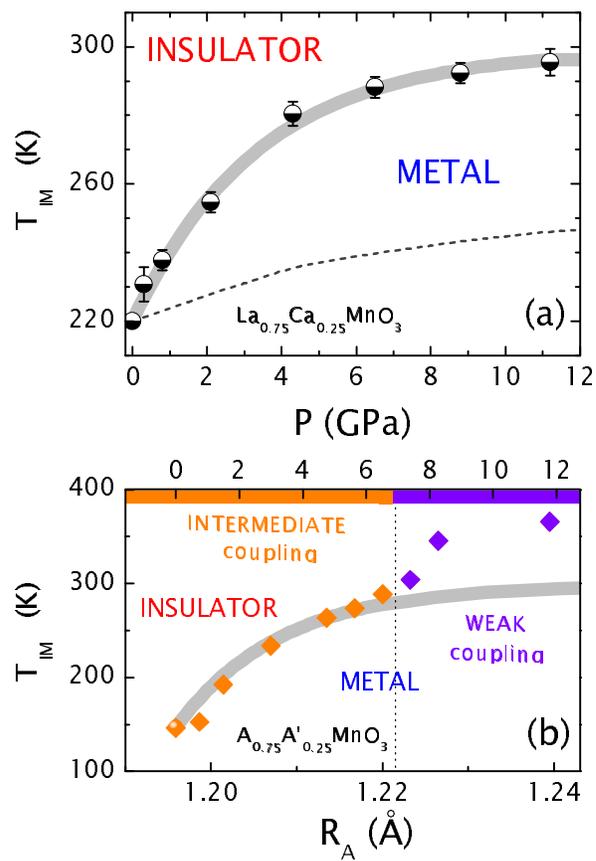
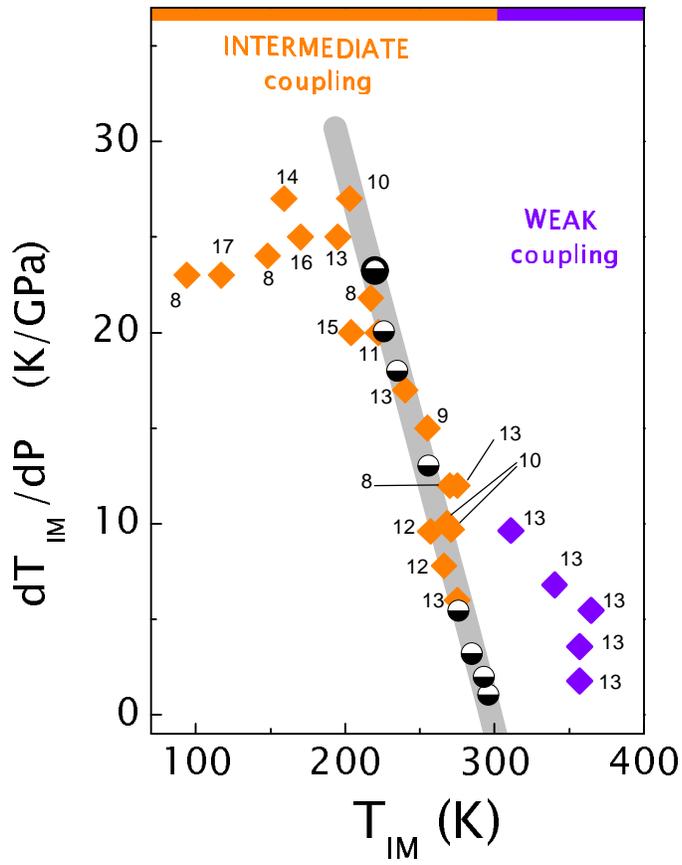
\end{document}